\documentclass[preprint,12pt]{elsarticle}

\usepackage{amssymb,amsmath,mathbbol,bm}

\journal{a journal}

\newtheorem{theorem}{Theorem}
\newtheorem{proposition}{Proposition}
\newtheorem{lemma}{Lemma}

\usepackage{xcolor}

\begin{document}

\begin{frontmatter}

\title{Bogolyubov's averaging theorem applied to the Kramers-Henneberger Hamiltonian}

\author[amu]{E. Floriani}

\address[amu]{Aix-Marseille Univ,  Universit\'e de Toulon, CNRS, CPT, Marseille, France}

\author[mpipks]{J. Dubois}

\address[mpipks]{Max Planck Institute for the Physics of Complex Systems, N\"{o}thnitzer Stra\ss{e} 38, Dresden, 01187, Germany}
            
\author[cnrs]{C. Chandre}

\address[cnrs]{CNRS, Aix-Marseille Universit\'e, Centrale  Marseille, I2M, Marseille, France}

\begin{abstract}
We apply Bogolyubov's averaging theorem to the motion of an electron of an atom driven by a linearly polarized laser field in the Kramers-Henneberger frame. We provide estimates of the differences between the original trajectories and the trajectories associated with the averaged system as a function of the parameters of the laser field and the region of phase space. We formulate a modified Bogolyubov averaging theorem based on the Hamiltonian properties of the system, and show that this version is better suited for these systems. From these estimates, we discuss the validity of the Kramers-Henneberger approximation.  
\end{abstract}

%%Research highlights
%\begin{highlights}
%\item Research highlight 1
%\item Research highlight 2
%\end{highlights}


\end{frontmatter}

\section{Introduction}
\label{sec:intro}

The motion of electrons in a combined Coulomb and time-dependent electric field displays a rather wide variety of behaviours which are easily understood only in two limits: First, when the electron is close to its ionic core, its motion is mainly driven by the Coulomb force, which corresponds to a case where the electric field acts as a small perturbation of the integrable case, therefore it evolves typically on invariant tori. Second, when the electron is far away from the ionic core, the electron is mostly driven by the electric field and it ionizes quickly. This second limit also corresponds to an integrable situation or a small perturbation of it. In between these two limits, the motion of the electron cannot be treated perturbatively, and typically displays features at different spatial and temporal scales. Remarkably, this is the most interesting situation dynamically and also for practical applications. Indeed, high harmonics of the driving field are generated by these trajectories which wander chaotically between the two integrable situations~\cite{vandeSand1999}. For example, the so-called recollisions, the keystone of strong-field atomic physics~\cite{Corkum1993,Schafer1993,Becker2008_ContP}, return to the ionic core after a pre-ionization. Another example is afforded by the ionization stabilization in strong fields~\cite{Grobe1991,Gavrila2002,Norman2015} where the ionization rate decreases as intensity is increased and a stable state is created as a result of the combination of the Coulomb potential and the time-dependent electric field.    

A natural way to understand some of the dynamical features experienced by the electron in a combined Coulomb and electric field is to perform averaging over the smaller temporal scales (see, e.g., Refs.~\cite{Dubois2018,Dubois2018_PRE}). It allows the reduction of phase space by focusing on more relevant time scales. An important milestone~\cite{Kramers1956,Henneberger1968} was achieved by Kramers and Henneberger with a change of coordinates adapted to the electric field and an averaging of the potential in these new coordinates. In this frame, the potential highlights a local minimum and therefore the possibility for a stable state, dressed by the field, far away from the ionic core. This has given rise to a quite extensive literature on the Kramers-Henneberger (KH) approximation~\cite{Breuer1992,Reed1993,Volkova1994,Popov1999,Forre2005,Ivanov2005,He2020} which led to the denomination of Kramers-Henneberger atom~\cite{Morales2011,Richter2013,Wei2017}. The importance of this reduction is undeniable. The main question we address here is how valid the approximation is, and if so, under which conditions. This question has been addressed in Refs.~\cite{Popov1999,Smirnova2000}. Classically, it amounts to estimating the distance between the trajectories of the original system with the ones of the averaged system. The averaging would be valid if this distance is small enough for at least the duration of the laser pulse (at least of the order of a few laser cycles). We follow Ref.~\cite{Smirnova2000} to derive rigorous estimates between the trajectories using Bogolyubov's averaging theorem. Bogolyubov's theorem is used to determine in which regions of phase space and for which region in parameter space an averaging method is valid for a given time. 
Here we show that this theorem, as stated in Refs.~\cite{Bogolyubov1961,Zhuravlev1988,Mitropolskii1997} cannot provide sufficiently accurate analytic estimates to fully address the question of the validity. We illustrate this point using a soft-Coulomb potential, which fully takes into account the long-range Coulomb interaction.

In order to provide estimates closer to the ionic core region, we formulate a modification of Bogolyubov's averaging theorem which is more adapted to Hamiltonian systems. The resulting estimates are shown to be much more accurate than the ones of the original formulation of Bogolyubov's theorem. They allow us to bring more elements on the validity of the KH approximation. This question cannot be simply answered by a consideration of the values of the parameters of the laser field only: 
the regions in phase space which are visited by the trajectories are paramount. 

Here, we consider the Hamiltonian of an electron in a combined ion potential and electric field in the dipole approximation
\begin{equation}
\label{eqn:H1}
H_{\rm e} ({\bf r}_{\rm e},{\bf p}_{\rm e},t) = \frac{{\bf p}_{\rm e}^2}{2} + V({\bf r}_{\rm e}) + {\bf r}_{\rm e}\cdot {\bf E}(t) \,,
\end{equation}
where $\mathbf{r}_{\rm e}$ is the position of the electron and $\mathbf{p}_{\rm e}$ its canonically conjugate momentum. The electron-ion potential $V (\mathbf{r}_{\rm e}) = - ( | \mathbf{r}_{\rm e} |^2+a^2)^{-1/2}$ is modeled as a soft-Coulomb potential.
The electric field is linearly polarized and is given by
$$
{\bf E}(t) =E_0 \cos(\omega t )\, \hat{\bf x} =-\frac{\partial {\bf A}}{\partial t} \,,
$$
where $\mathbf{A}(t)$ is the vector potential.

We move to the KH frame by performing the canonical change of coordinates
\begin{subequations}
\label{chvar2}
\begin{align}
    {\bf p} = {\bf p}_{\rm e} - {\bf A}(t) \,, \\
    {\bf r} = {\bf r}_{\rm e} - \frac{{\bf E}(t)}{\omega^2} \,.
\end{align}
\end{subequations}
In the KH frame, Hamiltonian~\eqref{eqn:H1} becomes
\begin{equation} 
\label{eqn:H2}
H({\bf r},{\bf p}, t) = \frac{{\bf p}^2}{2}+V\left({\bf r}+\frac{{\bf E}(t)}{\omega^2} \right) \,. 
\end{equation}
In the literature on KH frame, an approximation -coined Kramers-Henneberger approximation- is performed in addition to the change of coordinates.  
In this approximation, the above time-dependent Hamiltonian is averaged over one period of the laser field, and becomes 
\begin{equation} 
\label{E2}
\langle H \rangle ({\bf r}, {\bf p}) = \frac{{\bf p}^2}{2}+ V_{\rm KH}({\bf r}) \,,
\end{equation}
where
$$
V_{\rm KH} ({\bf r})=\frac{1}{T}\int_0^T V\left({\bf r}+\frac{{\bf E}(t)}{\omega^2} \right)\ {\rm d}t \,.
$$
The main gain is that $\langle H \rangle$ is now time-independent, and it now corresponds to a conserved quantity. As a consequence, its dynamics is more easily understood. However, a loss of dynamical information is associated with this approximation, since $\langle H \rangle$ is no longer conjugated to $H$.
The central question is how much information has been lost, moving from $H$ to $\langle H \rangle$, and overall, how good the KH approximation is. In order to bring elements to answer this question, we compare the trajectories of Hamiltonian~\eqref{eqn:H2} with the ones of Hamiltonian~\eqref{E2}. Equivalently, this brings some comparison between the trajectories
$\left( {\bf r}_{\rm e}(t),{\bf p}_{\rm e}(t) \right)$ of the original system~\eqref{eqn:H1} and the reconstructed trajectories $\left( \overline{{\bf r}}(t)+{\bf E}(t)/\omega^2,\overline{{\bf p}}(t)+{\bf A}(t)\right)$ where $\left( \overline{{\bf r}}(t),\overline{{\bf p}}(t) \right)$ are trajectories of Hamiltonian~\eqref{E2}. 

In Sec.~\ref{sec:bogo1}, we recall the formulation of Bogolyubov's averaging theorem and provide a concise proof. We provide and discuss the resulting estimates for a soft-Coulomb potential in one dimension. Next, in Sec.~\ref{sec:bogo2}, we modify Bogolyubov's theorem by adapting the estimates to a Hamiltonian flow. We compare and discuss the resulting estimates on a one-dimensional soft-Coulomb potential.  

\section{Bogolyubov's averaging theorem}
\label{sec:bogo1}

\subsection{Statement of the theorem}

We consider Bogolyubov's averaging theorem for comparing the trajectories of the original system with the ones of averaged system. This was the approach followed in Ref.~\cite{Smirnova2000} to assess the validity of the Kramers-Henneberger approximation.
We briefly recall the statement of the theorem and a brief proof for the periodic case (see Refs.~\cite{Bogolyubov1961,Zhuravlev1988,Mitropolskii1997} for more details). 
We consider the following equations of motion:
$$
\frac{ {\rm d} {\mathbf z}}{{\rm d}t} = \varepsilon\,  {\mathbf Z}(t,  {\mathbf z}, \varepsilon)\,,
$$
where $\varepsilon$ is a small positive parameter and $\mathbf{z}\in {\mathbb R}^n$ are the coordinates of the electron in phase space.
Two Cauchy problems are compared:
\begin{subequations} 
\label{eqmot}
\begin{align}
& \mbox{I} \, : \quad \frac{{\rm d} {\mathbf z}}{{\rm d}t} = \varepsilon\,  {\mathbf Z}(t,  {\mathbf z}, \varepsilon) \;, \quad\mbox{ with  }  {\mathbf z}(0) =  {\mathbf z}_0\,, \\
& \mbox{II}  \, : \quad \frac{{\rm d} {\mathbf u}}{{\rm d}t} = \varepsilon\, \langle {\mathbf Z}\rangle({\mathbf u}, \varepsilon)  \;, \quad\mbox{ with  }  {\mathbf u}(0) =  {\mathbf z}_0\,.
\end{align}
\end{subequations}
Here ${\mathbf Z}(t,  {\mathbf z}, \varepsilon)$ is defined in a domain ${\cal D}$ of ${\mathbf z}$. For $t\ge 0$ and $0\le\varepsilon\le\varepsilon_1$ the following conditions are assumed:
\begin{itemize}
\item[(a)]  
${\mathbf Z}(t,  {\mathbf z}, \varepsilon)$ is a $t$-measurable function for fixed ${\mathbf z}$ and $\varepsilon$, and the limit
\begin{equation} \label{defZav}
 \langle {\mathbf Z}\rangle({\mathbf z}, \varepsilon)  = \lim_{T\rightarrow\infty} \frac{1}{T} \int_0^T  {\mathbf Z}(t,  {\mathbf z}, \varepsilon)\, {\rm d}t \,,
\end{equation}
exists uniformly relative to both ${\mathbf z}\in {\cal D}$ and $\varepsilon \in [0,\varepsilon_1]$.
\item[(b)]
It is possible to find constants $M$ and $\lambda$ such that, for any  ${\mathbf z}, {\mathbf z'}$ in ${\cal D}$,
\begin{subequations}
\label{def_ML}
\begin{align} 
&\Vert {\mathbf Z} ({\mathbf z}, \varepsilon) \Vert \le M \,, \\
&\left\Vert {\mathbf Z}(t,  {\mathbf z}, \varepsilon) -  {\mathbf Z}(t,  {\mathbf z'}, \varepsilon) \right\Vert \le \lambda \Vert{\mathbf z} - {\mathbf z'}\Vert \,. \label{def_MLb}
\end{align}
\end{subequations}
\end{itemize}
We consider the case where ${\mathbf Z}(t,  {\mathbf z}, \varepsilon)$ is periodic in $t$, i.e., there exists $T$ such that ${\mathbf Z}(t,  {\mathbf z}, \varepsilon) = {\mathbf Z}(t+T,  {\mathbf z}, \varepsilon)$ for all $t\geq 0$ and ${\bf z} \in {\cal D}$.

\begin{theorem}[\cite{Bogolyubov1961,Zhuravlev1988}] 
If the solution ${\mathbf u}(t)$ to the initial Cauchy problem for system II is defined for $t\ge 0$ and belongs to the domain ${\cal D}$,
then, under Hypotheses (a) and (b), for any $L> 0$, the following inequality holds
\begin{equation} \label{thBog}
\Vert {\mathbf z}(t) - {\mathbf u}(t) \Vert \le  \varepsilon \,  \mathrm{e}^{\lambda L} \,T \left( L\lambda M + 2 M\right) \,,
\end{equation}
for all $0\le t<L/\varepsilon$ and $0\le\varepsilon<\varepsilon_1$.
\end{theorem}

{\it Proof:} From the obvious statement,
\begin{eqnarray*}
{\mathbf z}(t) - {\mathbf u}(t) &=& \varepsilon \int_0^t [ {\mathbf Z}(\tau,  {\mathbf z}(\tau), \varepsilon) -  \langle {\mathbf Z}\rangle({\mathbf u}(\tau), \varepsilon) \rangle ] {\rm d}\tau, \\
&=& \varepsilon \int_0^t [ {\mathbf Z}(\tau,  {\mathbf z}(\tau), \varepsilon) - {\mathbf Z}(\tau,  {\mathbf u}(\tau), \varepsilon) ]\, {\rm d}\tau \\
&& \qquad + \; \varepsilon \int_0^t [ {\mathbf Z}(\tau,  {\mathbf u}(\tau), \varepsilon) - \langle {\mathbf Z}\rangle({\mathbf u}(\tau), \varepsilon) \rangle ] \, {\rm d}\tau,
\end{eqnarray*}
where we have used the fact that ${\bf z}(0)-{\bf u}(0)={\bf 0}$, and from Eq.~\eqref{def_ML}, we deduce
$$
\Vert {\mathbf z}(t) - {\mathbf u}(t) \Vert \le \varepsilon \lambda \int_0^t \Vert {\mathbf z}(\tau) - {\mathbf u}(\tau) \Vert \, {\rm d}\tau + \sup_{t\in [0,L/\varepsilon]} \left\Vert \int_0^t  \varepsilon \, \Delta \mathbf {Z}(\tau,  {\mathbf u}(\tau), \varepsilon) \, {\rm d}\tau \right \Vert \,,
$$
where
$$
\Delta \mathbf{Z}(t,  {\mathbf u}(t), \varepsilon) = {\mathbf Z}(t,  {\mathbf u}(t), \varepsilon) - \langle {\mathbf Z}\rangle({\mathbf u}(t), \varepsilon) \,.
$$
Gr{\"o}nwall's lemma implies
$$
\left\Vert {\mathbf z}(t) - {\mathbf u}(t) \right\Vert \le \mathrm{e}^{\varepsilon\lambda t} \sup_{t\in [0,L/\varepsilon]} \left \Vert \int_0^t \varepsilon \, \Delta \mathbf {Z}(\tau,  {\mathbf u}(\tau), \varepsilon) \, {\rm d}\tau \right\Vert \,. 
$$
Therefore, the estimate on the error $\Vert {\mathbf z}(t) - {\mathbf u}(t) \Vert$ mainly depends on the maximum gradient of the flow, i.e., $\lambda$, and on the time integral of the fluctuating part of the flow computed at the solution of the averaged equations. In particular, the exponential term does not depend on the fact that the same initial conditions were used for both systems I and II.

In the periodic case, in order to estimate the integral of $\Delta \mathbf {Z}$, the interval $[0,t]$ is partitioned: For $n T\le t < (n+1)T$, i.e., $n=\lfloor t/T\rfloor$,
$$
\left\Vert \int_0^t \Delta \mathbf {Z}(\tau,  {\mathbf u}(\tau), \varepsilon) \, {\rm d}\tau \right\Vert \le  \sum_{i=1}^n \left\Vert \int_{(i-1)T}^{iT} \Delta \mathbf{Z}(\tau,  {\mathbf u}(\tau), \varepsilon) \, {\rm d}\tau \right\Vert + \left\Vert \int_{nT}^t \Delta \mathbf{Z}(\tau,  {\mathbf u}(\tau), \varepsilon) \, {\rm d}\tau \right\Vert\,.
$$
We denote the values of ${\mathbf u}$ at each period as ${\mathbf u}_i = {\mathbf u}((i-1)T)\,, \, i=1,\dots,n$. Given that 
$$
\int_{(i-1)T}^{iT} \Delta \mathbf{Z}(\tau,  {\mathbf u}_i, \varepsilon) \, {\rm d}\tau = 0\, ,
$$
which is obtained from the periodicity of the flow, we substitute this integral into the sum:
\begin{eqnarray}
\left\Vert \int_0^t \Delta \mathbf{Z}(\tau, {\mathbf u}(\tau), \varepsilon) \, {\rm d}\tau \right\Vert &\le & \sum_{i=1}^n  \int_{(i-1)T}^{iT} \Vert \Delta \mathbf{Z}(\tau,  {\mathbf u}(\tau), \varepsilon) - \Delta \mathbf{Z}(\tau,  {\mathbf u}_i, \varepsilon) \Vert {\rm d}\tau \nonumber \\
&& \qquad \qquad + \int_{nT}^t \Vert \Delta \mathbf{Z}(\tau,  {\mathbf u}(\tau), \varepsilon) \Vert \, {\rm d}\tau , \nonumber \\
& \le &  \sum_{i=1}^n 2 \lambda \int_{(i-1)T}^{iT} \Vert {\mathbf u}(\tau) - {\mathbf u}_i \Vert \, {\rm d}\tau + 2 M T\, , \label{eqn:eq3ln}
\end{eqnarray}
where the inequality $\Vert \langle \mathbf{Z} \rangle \Vert \leq M$ resulting from Eqs.~\eqref{defZav} and~\eqref{def_MLb} was used.
Now, we use the fact that for $\tau\in[(i-1)T,iT]$:
$$
{\mathbf u}(\tau) = {\mathbf u}_i + \int_{(i-1)T}^{\tau}  \varepsilon\, \langle {\mathbf Z}\rangle({\mathbf u}(s), \varepsilon) \, {\rm d}s\,,
$$
so that
\begin{equation}
\label{eqn:uui}
    \Vert {\mathbf u}(\tau) - {\mathbf u}_i \Vert \le \varepsilon M (\tau - (i-1)T) \,,
\end{equation}
which in turn, implies that
$$
\int_{(i-1)T}^{iT} \Vert {\mathbf u}(\tau) - {\mathbf u}_i \Vert \, {\rm d}\tau \le \varepsilon M \frac{T^2}{2} \,.
$$
Finally,
$$
\left\Vert \int_0^t \Delta \mathbf{Z}(\tau, {\mathbf u}(\tau), \varepsilon) {\rm d}\tau \right\Vert \le  2\lambda n \,\varepsilon M \, \frac{T^2}{2} + 2 M T \leq T \left(\varepsilon t \lambda M  + 2 M \right) \,.
$$
Since $\varepsilon t < L$, Eq.~\eqref{thBog} follows.

\bigskip

Equation~\eqref{thBog} is the original formulation of Bogolyubov's averaging theorem. However, when examining the proof, it can be seen that a sharper version of it can be given, essentially by introducing specific bounds for the flows $\langle {\mathbf Z}\rangle$ and $\Delta{\mathbf Z}$. We show below that this formulation improves the estimates of the discrepancy between the original and the averaged trajectories, and in fact, is better suited to a system with energy conservation. The modified statement reads as follows:

\begin{proposition} 
If there are constants $\overline{M}$, $\lambda$, $M_{\delta}$ and $\lambda_{\delta}$ such that, for any  ${\mathbf z}, {\mathbf z'}$ in ${\cal D}$, and for all $t\ge 0$ 
\begin{subequations}
\label{def_MLd}
\begin{align} 
&\Vert \langle {\mathbf Z} \rangle({\mathbf z}, \varepsilon) \Vert \le \overline{M} \,, \\
&\left\Vert {\mathbf Z}(t,  {\mathbf z}, \varepsilon) -  {\mathbf Z}(t,  {\mathbf z'}, \varepsilon) \right\Vert \le \lambda \Vert{\mathbf z} - {\mathbf z'}\Vert \,, \\
&\Vert \Delta \mathbf{Z}(t,   {\mathbf z}, \varepsilon) \Vert \le M_{\delta} \,, \\
&\left\Vert \Delta \mathbf{Z}(t,  {\mathbf z}, \varepsilon) -  \Delta \mathbf{Z}(t,  {\mathbf z'}, \varepsilon) \right\Vert \le \lambda_{\delta}  \Vert{\mathbf z} - {\mathbf z'}\Vert \, ,
\end{align}
\end{subequations}
where $\Delta \mathbf{Z}= {\bf Z}-\langle {\bf Z}\rangle$, then the statement \eqref{thBog} is modified as follows:
\begin{equation} \label{thBogd}
\Vert {\mathbf z}(t) - {\mathbf u}(t) \Vert \le  \varepsilon \,  \mathrm{e}^{\lambda L} \,T \left( \frac{L}{2} \lambda_{\delta} \overline{M} + M_{\delta} \right) \,.
\end{equation}
\end{proposition}

We notice that Proposition 1 always provides better estimates than Theorem 1, since
$$
\varepsilon \,  \mathrm{e}^{\lambda L} \,T \left( \frac{L}{2} \lambda_{\delta} \overline{M} + M_{\delta} \right) \leq \varepsilon \,  \mathrm{e}^{\lambda L} \,T \left( L \lambda M + 2 M \right).
$$
The proof of this proposition is identical to the proof of Theorem 1. We notice that the constant $\overline{M}$ is the maximum value of the averaged flow, while $M$ is the maximum value of the flow itself in Theorem 1.

\subsection{Application to a one-dimensional soft-Coulomb potential} \label{sec_SC1d}
We consider a soft-Coulomb potential in one dimension. Hamiltonian~\eqref{eqn:H2} becomes
$$
H(x,p,t) = \frac{p^2}{2} + V\left(x+\frac{E(t)}{\omega^2} \right) \,,
$$
where 
$$
V\left(x+\frac{E(t)}{\omega^2} \right) = - \frac{1}{\sqrt{ (x + q \cos(\omega t))^2 + a^2 }} \,,
$$
with $q=E_0/\omega^2$ the quiver radius.
 
\noindent {\em Rescaling procedure :} 
We cast the equations of motion associated with Hamiltonian $H$ in the form \eqref{eqmot} in order to apply Theorem 1. We define  $\phi = \omega t$ as the rescaled time/evolution parameter, and rescale position and momentum as
$$
x =C_x\, \chi \;, \quad\quad p = C_p\, \xi \,. 
$$
Inserting these rescalings, the equations of motion become
$$
\frac{{\rm d} \chi}{{\rm d}\phi} =  \frac{C_p}{\omega\, C_x}\, \xi \;, \quad\quad \frac{{\rm d}\xi}{{\rm d}\phi} = - \frac{1}{\omega\, C_p C_x^2}\, W'(\chi + (q/C_x)\cos\phi) \,,
$$
where $W$ is the rescaled potential:
$$
V(x+q \cos\omega t) = \frac{1}{C_x}\, W\left(\chi + (q/C_x) \cos\phi \right) \,.
$$
So, the factors $C_x, C_p$ and the small parameter $\varepsilon$ of Bogolyubov's theorem must satisfy the relation
$$
\frac{C_p}{\omega\, C_x} = \frac{1}{\omega\, C_p C_x^2} = \varepsilon \, .
$$
The choice of $C_x$ is guided by the fact that the rescaling of time and space is ruled by the external field, which governs the fast dynamics. That is why we choose $C_x$ to be the quiver radius $q$. We then have
\begin{equation} \label{rescaling}
\varepsilon = \frac{1}{\omega q^{3/2}} = \frac{\omega^2}{E_0^{3/2}} \;, \quad\quad 
\phi = \omega t \;, \quad\quad \chi = \frac{x}{q} \;, \quad\quad \xi = \sqrt q \,p \, , 
\end{equation}
where $\chi$ and $\xi$ are the rescaled position and momentum, respectively.
The equations of motion in the rescaled variables are:
$$
\frac{{\rm d} \chi}{{\rm d}\phi} = \varepsilon \, \xi \;, \quad\quad \frac{{\rm d}\xi}{{\rm d}\phi} = - \varepsilon \, W'(\chi + \cos\phi) \,,
$$
where $\mathbf{z} = (\chi , \xi)$ in the notations of Theorem 1 and
$$
W(\chi) = - \frac{1}{\sqrt{ \chi^2 + \alpha^2 }} \,, 
$$
where $\alpha = a/q$ is the rescaled softening parameter. 

Using the KH approximation, the rescaled KH potential is denoted $W_{\rm KH}$ and reads
$$
W_{\rm KH}(\chi) = - \frac{1}{2\pi} \int_0^{2\pi} \frac{{\rm d}\phi}{\sqrt{ (\chi + \cos\phi)^2 + \alpha^2}} = - \left\langle \frac{1}{\sqrt{ (\chi + \cos\phi)^2 + \alpha^2}} \right\rangle .
$$
Therefore the rescaled potential is related to the unrescaled potential by $W_{\rm KH}(\chi) = q V_{\rm KH}(q\chi)$.
The averaged trajectory $ {\mathbf u} = (\overline{\chi} , \overline{\xi})$ obeys the equations:
$$
\frac{{\rm d} \overline{\chi}}{{\rm d}\phi} = \varepsilon \, \overline{\xi} \;, \quad\quad \frac{{\rm d}\overline{\xi}}{{\rm d}\phi} =  - \varepsilon \, W_{\rm KH}'(\overline{\chi}).
$$
The flows involved in Theorem 1 are given by
\begin{eqnarray*}
&& \mathbf {Z}(\phi, {\mathbf z}) = \left( \xi \; , - W'\left(\chi + \cos\phi\right) \right) \;, \\
&& \langle \mathbf{Z} \rangle({\mathbf u}) = \left( \overline{\xi} \;,  - W_{\rm KH}'(\overline{\chi}) \right).
\end{eqnarray*}
In what follows, we use the 1-norm:
$$
\Vert{\mathbf z} \Vert = | \chi | + | \xi |. 
$$
Other choices would give similar results, thanks to the equivalence of norms in finite dimension.

The constants $M$, $\lambda$, $\overline{M}$, $M_\delta$ and $\lambda_\delta$ appearing in Theorem 1 and Proposition 1 are expressed as functions of the bounds for the absolute values of $\xi$, $W$, $W_{\rm KH}$ and their derivatives. We denote these by
\begin{eqnarray*} 
| \xi |  \le B_{p} \;, \qquad  | W(\chi+\cos\phi) |  \le B_{W} \;, \qquad \left|  W^{(n)}(\chi+\cos\phi) \right| \le B_{W,n} \;, \\ \nonumber
| W_{\rm KH}(\chi) |  \le B_{\langle W\rangle} \;, \qquad \left|  W_{\rm KH}^{(n)}(\chi) \right| \le B_{\langle W\rangle,n} \,.
\end{eqnarray*}
These bounds have to hold for any $\phi\ge 0$ and will depend on the region $(\chi,\xi)$ of phase space where the trajectory evolves. We have then
$$
M = \overline{M} =  B_p + B_{W,1}  \,, 
\quad \lambda = \max \left( 1, B_{W,2} \right) \,,
\quad M_\delta =  B_{W,1}  \,, 
\quad \lambda_\delta = B_{W,2} \,. 
$$
From Theorem 1, for all $0 \le \phi \le L/\varepsilon$, we have:
$$
| \chi(\phi) - \overline{\chi}(\phi)  | + | \xi(\phi)  - \overline{\xi}(\phi) | \le  \varepsilon \,  \mathrm{e}^{\lambda L} \,2\pi \left( L \lambda M + 2 M \right) \,,
$$
and, from Proposition 1,
$$
| \chi(\phi) - \overline{\chi}(\phi)  | + | \xi(\phi)  - \overline{\xi}(\phi) | \le   \varepsilon \,  \mathrm{e}^{\lambda L} \,2\pi \left( \frac{L}{2} \lambda_{\delta} \overline{M} + M_{\delta} \right) \,.
$$
To deduce an estimate for the original variables $x_{\rm e}$, $p_{\rm e}$ and $t$, Eqs.~\eqref{chvar2} and \eqref{rescaling} are used, i.e., 
$$
x_{\rm e}(t) = q [ \chi(\omega t) + \cos(\omega t) ] \;, \quad\quad  p_{\rm e}(t) = \frac{1}{\sqrt q} \left[ \xi(\omega t) - \frac{1}{\varepsilon} \,\sin(\omega t) \right] \,,
$$
together with the analogous relations for the reconstructed trajectories of the averaged system (ruled by the time-independent Hamiltonian~\eqref{E2}), which are denoted by $\overline{x}_{\rm e}(t), \overline{p}_{\rm e}(t)$:
$$
\overline{x}_{\rm e}(t) = q [ \overline{\chi}(\omega t) + \cos(\omega t) ] \;, \quad\quad  \overline{p}_{\rm e}(t) = \frac{1}{\sqrt q} \left[ \overline{\xi}(\omega t) - \frac{1}{\varepsilon} \,\sin(\omega t) \right] \,.
$$
We notice that, given these expressions, we have $\vert x_{\rm e}(t) -\overline{x}_{\rm e}(t)\vert = \vert x(t)-\overline{x}(t)\vert$ and $\vert p_{\rm e}(t) -\overline{p}_{\rm e}(t)\vert = \vert p(t)-\overline{p}(t)\vert$. 

 We distinguish the case where $\chi$ takes values on the whole real line, and the case where $\chi$ stays larger than a prescribed value. In each case, we estimate the bounds $B$s, using energy conservation on averaged trajectories
$$
E_{\rm KH} = \frac{1}{q} \left[ \frac{\xi(0)^2}{2} + W_{\rm KH}(\chi(0)) \right] = \frac{1}{q} \left[ \frac{\overline{\xi}(\phi)^2}{2} + W_{\rm KH}(\overline{\chi}(\phi)) \right]
$$
for all $\phi$.
We notice that the bounds $M$, $\overline{M}$, $M_\delta$ and $\lambda_\delta$ are obtained from the averaged trajectory ${\bf u}(t)$, which allows the use of the energy conservation for their estimates (see, e.g., Eqs.~\eqref{eqn:eq3ln}-\eqref{eqn:uui}). 

\subsubsection{Case $\chi \in {\mathbb R}$} \label{SC_R2}

If we consider the whole domain ${\mathbb R}^{2}$ for $(\chi,\xi)$, then 
\begin{equation} 
\label{BWsoft2}
B_{W} = \vert W(0)\vert = \frac{1}{\alpha} \;, \quad B_{W,1} = \left\vert W'\left( \frac{\alpha}{\sqrt 2} \right) \right\vert = \frac{2}{3\sqrt{3}\, \alpha^2 } \;, \quad B_{W,2} = \vert W''(0)\vert = \frac{1}{\alpha^3} \,,
\end{equation}
and since we do not have an explicit expression for $W_{\rm KH}$ we take $B_{\langle W\rangle,n} = B_{W,n}$. For bound states ($E_{\rm KH} < 0$), energy conservation on averaged trajectories implies that
$$
| \overline{\xi} | \le \sqrt{ 2 \left( q E_{\rm KH}+ B_{W} \right) } \le \sqrt{ 2 B_{W} } = B_p \,.
$$

\noindent {\it In sum: } Theorem 1 gives, for all $0\le t<t_{\rm max}$:
$$
\frac{1}{q} \, \left| x_{\rm e}(t) - \overline{x}_{\rm e}(t) \right| + \sqrt q \, \left| p_{\rm e}(t) - \overline{p}_{\rm e}(t) \right| \le \frac{T}{q^{3/2}} \, \mathrm{e}^{\lambda\,  t_{\rm max}\, q^{-3/2} } \left( \frac{t_{\rm max} \lambda M }{q^{3/2}} + 2M \right) \,,
$$
where
$$
M = \frac{2}{3\sqrt{3}\, \alpha^2 } + \sqrt{ \frac{2}{\alpha} }  \;, \qquad \lambda = \max\left( 1, \frac{1}{\alpha^3}  \right) \,.
$$
If Proposition 1 is used, we have
$$
\frac{1}{q} \, \left| x_{\rm e}(t) - \overline{x}_{\rm e}(t) \right| + \sqrt q \, \left| p_{\rm e}(t) - \overline{p}_{\rm e}(t) \right| \le \frac{T}{q^{3/2}} \, \mathrm{e}^{\lambda\,  t_{\rm max}\, q^{-3/2} } \left( \frac{t_{\rm max} \lambda_{\delta} M }{2\, q^{3/2}} + M_{\delta} \right) \,,
$$
with
$$
M_{\delta} = \frac{2}{3\sqrt{3}\, \alpha^2 } \;, \qquad \lambda_{\delta} = \frac{1}{\alpha^3} \,.
$$

\bigskip

For typical values of laser intensity $I = 5\times 10^{15}$ W cm$^{-2}$, frequency $\omega = 0.0584$, $a=1$ and $t_{\rm max} = 5$ laser cycles, we have $M \simeq M_\delta \simeq 4.7\times 10^3$, $\lambda = \lambda_\delta \simeq 1.4 \times 10^6$. So, the preceding bounds are mostly useless when evaluated at realistic values of the parameters. 
This is due to the fact that the electron is allowed to go arbitrarily close to the ionic core: the value of the bounds \eqref{BWsoft2} is reached when the position $q (\chi+\cos\phi)$ of the electron is zero or close to zero. \\ 
{\em Remark:} The same bounds apply to the case where the trajectory of the averaged system lies near the local minimum $\chi_*$ of the KH potential, i.e., to the case $1-\mu \le |\chi | \le 1$, with $\mu\in [0,1]$. This is due to the fact that the absolute value of the argument $\chi+\cos\phi$ of the potential $W$ takes all the values from $0$ to $2-\mu$ as $\phi$ varies, so that the absolute maxima of $\vert W\vert$ and $\vert W''\vert$ given by Eq.~\eqref{BWsoft2} are attained. In particular, the exponential behavior of the estimate is mostly identical for $\chi\in {\mathbb R}$ and for $\vert\chi\vert \in [1-\mu, 1]$.

\subsubsection{Case $|\chi | \ge 1+\mu$} \label{SC_Dmu}
We now consider the domain ${\mathcal D}_{\mu}$ defined by $|\chi | \ge 1+\mu$ (and arbitrary $\xi$) with $\mu$ a fixed positive number. The bounds are now: 
\begin{eqnarray*}
&& B_{W} = \vert W(\mu)\vert = \frac{1}{\sqrt{\mu^2 + \alpha^2}}  \quad \mbox{for any} \quad \mu>0 \,,\\
&& B_{W,1} = | W'(\mu) | = \frac{\mu}{ \left( \mu^2 + \alpha^2 \right)^{3/2}}  \quad \mbox{if} \quad  \mu\ge \frac{\alpha}{\sqrt 2} \,, \\
&& B_{W,2} = | W''(\mu) | =  \frac{2 \mu^2 - \alpha^2}{ \left( \mu^2 + \alpha^2 \right)^{5/2}}  \quad\mbox{if} \quad  \mu\ge \sqrt{ \frac{3}{2} }\, \alpha \,.
\end{eqnarray*}

\noindent {\it In sum: } For $x_{\rm e}$ and $\overline{x}_{\rm e}$ satisfying the condition $| x_{\rm e}(t) | , | \overline{x}_{\rm e}(t) | \ge \mu q$ and $\mu \ge \alpha \sqrt{ 3/2 }$, Theorem 1 gives, for all $0\le t<t_{\rm max}$:
\begin{equation}
\label{eqn:sigma2}
\frac{1}{q} \, \left| x_{\rm e}(t) - \overline{x}_{\rm e}(t) \right| + \sqrt q \, \left| p_{\rm e}(t) - \overline{p}_{\rm e}(t) \right| \le \frac{T}{q^{3/2}} \, \mathrm{e}^{\lambda\, t_{\rm max}\, q^{-3/2} } \left( \frac{t_{\rm max} \lambda M }{q^{3/2}} + 2 M \right) \,,
\end{equation}
where
$$
M = \frac{\mu}{ \left( \mu^2 + \alpha^2 \right)^{3/2}} + \frac{\sqrt{2}}{(\mu^2 + \alpha^2)^{1/4}}
 \;, \qquad\ \lambda = \max\left( 1, \frac{2 \mu^2 - \alpha^2}{ \left( \mu^2 + \alpha^2 \right)^{5/2}} \right) \,.
$$
If Proposition 1 is used, we have
\begin{equation}
\label{eqn:sigmad2}
\frac{1}{q} \, \left| x_{\rm e}(t) - \overline{x}_{\rm e}(t) \right| + \sqrt q \, \left| p_{\rm e}(t) - \overline{p}_{\rm e}(t) \right| \le \frac{T}{q^{3/2}} \, \mathrm{e}^{\lambda\,  t_{\rm max}\, q^{-3/2} } \left( \frac{t_{\rm max} \lambda_{\delta} M }{2\, q^{3/2}} + M_{\delta} \right) \,,
\end{equation}
with
$$
M_{\delta} = \frac{\mu}{ \left( \mu^2 + \alpha^2 \right)^{3/2}} \;, \qquad \lambda_{\delta} = \frac{2 \mu^2 - \alpha^2}{ \left( \mu^2 + \alpha^2 \right)^{5/2}} \,.
$$

\begin{figure}
    \centering
    \includegraphics[width=0.8\textwidth]{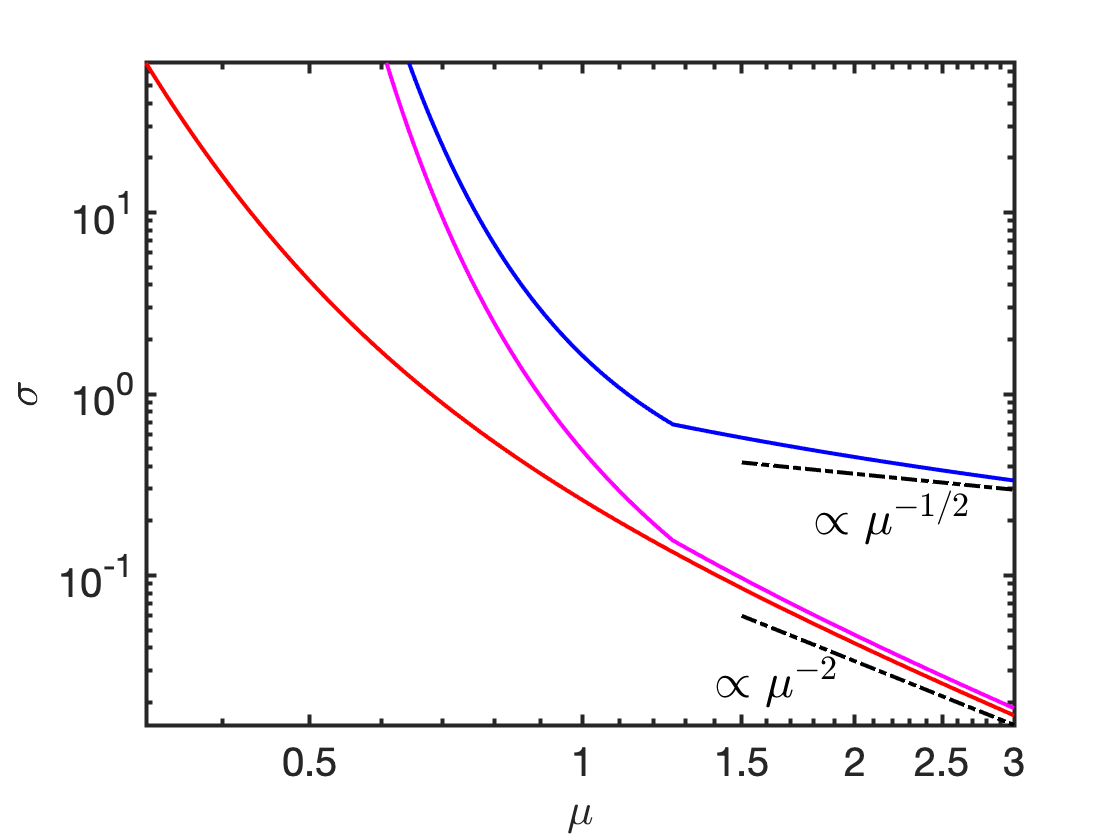}
    \caption{
Loglog plot of $\sigma =  \left| x_{\rm e}(t) - \overline{x}_{\rm e}(t) \right|/ q + \sqrt q \, \left| p_{\rm e}(t) - \overline{p}_{\rm e}(t) \right|$ as a function of $\mu$ at $I = 5\times 10^{15}$ W cm$^{-2}$, $\omega = 0.0584$, $a=1$ and $t_{\rm max} = 5$ laser cycles. Blue line: from Theorem 1, Eq.~\eqref{eqn:sigma2}. Magenta line: from Proposition 1, Eq.~\eqref{eqn:sigmad2}. Red line: from Theorem 2, Eq.~\eqref{eqn:sigma2H}.
}
    \label{fig:tradi}
\end{figure}

In Fig.~\ref{fig:tradi}, we represent the right-hand side of the estimates~\eqref{eqn:sigma2}-\eqref{eqn:sigmad2} as a function of $\mu$ (respectively, blue and magenta curves). We notice that the estimate given by Eq.~\eqref{eqn:sigmad2} is much more accurate for large values of $\mu$, i.e., when the trajectory is far away from the core (of the order of several quiver radii). Close to the ionic core, both estimates are equivalently inaccurate.

\subsection{Discussion} \label{sec_disc}

{\em Is the Kramers-Henneberger reduction valid?} In order to answer this question, Bogolyubov's averaging theorem and a variant of it were used. In the estimates presented in this section, the most damaging term is the exponential term, ${\rm e}^{\lambda t_{\rm max} q^{-3/2}}$ which can be rewritten as ${\rm e}^{\varepsilon \lambda \omega t_{\rm max} }$. This term is present in both Theorem 1 and Proposition 1 with the same exponent $\lambda$. This exponential term becomes ${\rm e}^{2\pi \varepsilon \lambda n }$, where $n$ is the duration of the laser pulse in laser cycles. This term remains of order one if $\varepsilon \lambda$ is much smaller than 1. The best-case scenario is when the electron is relatively far away from the ionic core, where $\lambda$ is of order 1 (notice that it is even larger close to the ionic core). This leads to an upper bound for $\varepsilon$ which translates into a lower bound for $E_0$:
$$
E_0\gg \omega^{4/3} (2\pi n)^{2/3}.
$$
For an infrared laser field with $\omega=0.0584$ with a duration of five laser cycles, this means an intensity greater than $1.8\times 10^{15}$ W cm$^{-2}$, which is much larger than what is typically considered for KH simulations. Following Ref.~\cite{Smirnova2000}, an alternative way to make this exponential term small is to reduce the size of the pulse, namely $n$, and considering
$$
n\lesssim \frac{1}{2\pi \lambda \sqrt{\varepsilon} } \,.
$$
For $\omega=0.0584$ and an intensity of $5\times 10^{15}$ W cm$^{-2}$, this gives $n$ less than approximately one laser cycle, which is not realistic (and it is even worse for smaller intensities). Therefore the only way to reduce the effects of the exponential term is to have parameters of the field so that $\varepsilon$ is typically smaller than 1/10, and being relatively far away from the ionic core, typically more than a couple of quiver radii. 

{\em Does this mean that the Kramers-Henneberger reduction is invalid? } Of course, not. This just means that estimates following Bogolyubov's theorem are not suited to address this question, at least at a distance closer than one quiver radius of the ionic core, since they neglect the true dynamics in phase space, chaotic in nature close to the bounded region and rather regular in large portions of phase space away from it. More importantly, the estimates are very general, and not tailored to Hamiltonian systems. To illustrate this point, we consider the following simple example for ${\bf z}=(x,p)$ and Hamiltonian $H(t,{\bf z},\varepsilon)=\varepsilon (p^2/2 -x \cos t)$:
\begin{subequations}
\label{sflow1}
\begin{align}
& \mathbf{Z} (t,\mathbf{z},\varepsilon)= \varepsilon (p \; , \cos t)\, , \\
& \langle \mathbf{Z} \rangle (\mathbf{z},\varepsilon)= \varepsilon (p \; , 0)\, .
\end{align}
\end{subequations}
The constant $\lambda$ in Theorem 1 and Proposition 1 is equal to 1, so that
$$
\Vert {\bf z}(t)-{\bf u}(t)\Vert \leq 2\pi \varepsilon  M (\varepsilon t_{\rm max} + 2) {\rm e}^{\varepsilon t_{\rm max}} \,, 
$$
for all $\vert p\vert \leq M-1$ and $t\leq t_{\rm max}$. However, in this case, the Hamiltonian flow~\eqref{sflow1} is easily integrated, and the estimate computed exactly:
$$
\Vert {\bf z}(t)-{\bf u}(t)\Vert = \varepsilon \vert \sin t\vert  + \varepsilon^2 \vert 1-\cos t \vert \leq  \varepsilon (1+2\varepsilon) \,,
$$
for all times. Instead of an exponential growth of the distance, this distance remains bounded in time. Therefore, we see on one of the simplest examples that estimates following Bogolyubov's theorem are far from being optimal. This is expected since these estimates are obtained with very general assumptions on the flow regardless of, e.g., its geometrical structure. 

Below, we modify Bogolyubov's theorem so that the estimates are better adapted to Hamiltonian systems. 

\section{Modification of Bogolyubov's averaging theorem}
\label{sec:bogo2}

\subsection{Statement of the theorem}

First, we separate the variables into positions and momenta:
$$
{\mathbf z} = ({\mathbf r} \, , {\mathbf p}) \,.
$$
The equations of motion are of the form
$$
\frac{{\rm d} {\mathbf z}}{{\rm d}t} = \varepsilon\, \left(\, \nabla_{\mathbf p} H \, , - \nabla_{\mathbf r} H \,\right) = \varepsilon\,  {\mathbf Z}(t,  {\mathbf z}, \varepsilon)\,.
$$
The trajectory ${\mathbf z}(t)$ is compared to the trajectory ${\mathbf u}(t) = (\overline{\mathbf r}(t) \, , \overline{\mathbf p}(t))$ of the averaged system
$$
\frac{{\rm d} {\mathbf u}}{{\rm d}t} = \varepsilon\, \left(\, \nabla_{{\mathbf p}} \langle H\rangle \, , - \nabla_{{\mathbf r}} \langle H\rangle \,\right) = \varepsilon\,  \langle {\mathbf Z}\rangle ({\mathbf u}, \varepsilon)\,.
$$
Following Bogolyubov's theorem, we consider that the two trajectories have the same initial condition ${\mathbf z}(0) = {\mathbf u}(0) = {\mathbf z}_0$. We estimate the differences in the positions and in momenta, i.e., the vector
$$
\begin{pmatrix}
\Vert {\mathbf r}(t) - \overline{\mathbf r}(t)\Vert \\
\Vert {\mathbf p}(t) - \overline{\mathbf p}(t)\Vert 
\end{pmatrix} = \sum_{i=1}^d
\begin{pmatrix}
| r_i(t) - \overline{r}_i(t) | \\
| p_i(t) - \overline{p}_i(t) |
\end{pmatrix}\,,
$$
where $d$ is the dimension of configuration space.

Two Cauchy problems are compared:
\begin{eqnarray*}
&& \mbox{I} \, : \; \frac{{\rm d} {\mathbf r}}{{\rm d}t} = \varepsilon\,   \nabla_{\mathbf p} H(t,  {\mathbf r},  {\mathbf p},\varepsilon) \;, \; \frac{{\rm d} {\mathbf p}}{{\rm d}t} = - \varepsilon\,   \nabla_{\mathbf r} H(t,  {\mathbf r},  {\mathbf p},\varepsilon)\,, \\
&& \mbox{II} \, : \; \frac{{\rm d} \overline{\mathbf r}}{{\rm d}t} = \varepsilon\,   \nabla_{{\mathbf p}} \langle H\rangle(\overline{\mathbf r},  \overline{\mathbf p},\varepsilon) \;, \; \frac{{\rm d} \overline{\mathbf p}}{{\rm d}t} = - \varepsilon\,   \nabla_{{\mathbf r}} \langle H\rangle(\overline{\mathbf r},  \overline{\mathbf p},\varepsilon) \,, 
\end{eqnarray*}
with ${\mathbf r}(0) =  \overline{\mathbf r}(0) ={\mathbf r}_0$ and ${\mathbf p}(0) =  \overline{\mathbf p}(0) ={\mathbf p}_0$.
Here $H(t,  {\mathbf r},  {\mathbf p},\varepsilon)$ is defined in a domain ${\cal D}$ of $({\mathbf r}, {\mathbf p})$, and is a periodic function of $t$, of period $T$. 

We suppose that, for $t \ge 0$ and $0\le\varepsilon < \varepsilon_1$, it is possible to define two-dimensional vectors
\begin{equation} \label{def_Mvect}
\overline{{\mathbf M}} = 
\begin{pmatrix}
M_p \\ M_r
\end{pmatrix} \,, \qquad {\mathbf M_\delta} = 
\begin{pmatrix}
M^\delta_p \\ M^\delta_r
\end{pmatrix} \,,
\end{equation}
and $2\times 2$ matrices
\begin{equation} \label{def_Lmatr}
{\mathbb \Lambda} =
\begin{pmatrix}
\lambda_{rp} & \lambda_{pp} \\ 
\lambda_{rr} & \lambda_{rp}
\end{pmatrix} \,, \qquad {\mathbb \Lambda}_\delta = 
\begin{pmatrix}
\lambda^\delta_{rp} & \lambda^\delta_{pp} \\ 
\lambda^\delta_{rr} & \lambda^\delta_{rp}
\end{pmatrix} \,,
\end{equation}
such that the following conditions hold 
for $t\ge 0$, for any ${\mathbf z} = ({\mathbf r} \, , {\mathbf p})$ in ${\cal D}$:
\begin{subequations}
\label{def_ML_H}
\begin{align} 
&\max_{1\le i\le d} |{\mathbf \nabla}_{ r_i} \langle H \rangle | \le \frac{M_r}{d}\,, \qquad \max_{1\le i\le d} |{\mathbf \nabla}_{ p_i} \langle H \rangle | \le \frac{M_p}{d} \,, \\
&\max_{1\le i\le d} |{\mathbf \nabla}_{r_i} \Delta H  | \le \frac{M^\delta_r}{d}\,, \qquad \max_{1\le i\le d} |{\mathbf \nabla}_{p_i} \Delta H  | \le \frac{M^\delta_p}{d} \,, \\
&\max_{1\le i,j\le d} |{\mathbf \nabla}_{r_i}{\mathbf \nabla}_{r_j} H | \le \frac{\lambda_{rr}}{d}\,, \; \max_{1\le i,j\le d} |{\mathbf \nabla}_{p_i}{\mathbf \nabla}_{p_j} H | \le \frac{\lambda_{pp}}{d} \,, \;
\max_{1\le i,j\le d} |{\mathbf \nabla}_{p_i}{\mathbf \nabla}_{r_j} H  | \le \frac{\lambda_{rp}}{d}\,,\\
&\max_{1\le i,j\le d} |{\mathbf \nabla}_{r_i}{\mathbf \nabla}_{r_j} \Delta H  | \le \frac{\lambda^\delta_{rr}}{d} \,, \; \max_{1\le i,j\le d} |{\mathbf \nabla}_{p_i}{\mathbf \nabla}_{p_j} \Delta H  | \le \frac{\lambda^\delta_{pp}}{d} \,, \;
\max_{1\le i,j\le d} |{\mathbf \nabla}_{p_i}{\mathbf \nabla}_{r_j} \Delta H  | \le \frac{\lambda^\delta_{rp}}{d} \,,
\end{align}
\end{subequations}
\color{black}
where
$$
\Delta H(t,  {\mathbf z}, \varepsilon) = H(t,  {\mathbf z}, \varepsilon) - \langle H\rangle({\mathbf z}, \varepsilon) \,.
$$
{\em Remark:} If we denote by ${\mathbb A}$, ${\mathbb A}_\delta$ the Hessian matrices of $H$, $\Delta H$, and ${\mathbb J}_{2d}$ the symplectic matrix:
$$
A_{ij} = {\mathbf \nabla}_{z_i}{\mathbf \nabla}_{z_j} H \,, \qquad
A_{ij}^\delta = {\mathbf \nabla}_{z_i}{\mathbf \nabla}_{z_j} \Delta H \,, \qquad
{\mathbb J}_{2d} =
\begin{pmatrix}
0 & \mathbb{I}_d \\ -\mathbb{I}_d & 0
\end{pmatrix} \,,
$$
where ${\mathbb I}_d$ is the $d\times d$ identity matrix, we see that $\overline{{\mathbf M}}$, ${\mathbf M_\delta}$ give bounds for the absolute value of the vectors ${\mathbb J}_{2d} \nabla_{\mathbf z} \langle H \rangle$, ${\mathbb J}_{2d} \nabla_{\mathbf z} \Delta H$, while ${\mathbb \Lambda}$, ${\mathbb \Lambda}_\delta$ give bounds for the absolute value of ${\mathbb J} {\mathbb A}$, ${\mathbb J} {\mathbb A}_\delta$.

\begin{theorem} 
If the solution $(\overline{\mathbf r}(t) \, , \overline{\mathbf p}(t))$ to the initial Cauchy problem for system II is defined for $t\ge 0$ and belongs to the domain ${\cal D}$, then for any $L>0$, the following inequality holds
\begin{eqnarray} \label{thBog_H}
\begin{pmatrix}
\Vert {\mathbf r}(t) - \overline{\mathbf r}(t)\Vert \\
\Vert {\mathbf p}(t) - \overline{\mathbf p}(t)\Vert 
\end{pmatrix} \le 
{\rm e}^{ {\mathbb \Lambda} L} \varepsilon T \left[\frac{L}{2} {\mathbb \Lambda}_\delta \overline{{\mathbf M}} + {\mathbf M_\delta} \right] \,,
\end{eqnarray}
for all $0\le t<L/\varepsilon$ and $0\le\varepsilon<\varepsilon_1$, with $\overline{{\mathbf M}}$, ${\mathbf M_\delta}$, ${\mathbb \Lambda}$ and ${\mathbb \Lambda}_\delta$ defined by Eqs.~(\ref{def_Mvect}), (\ref{def_Lmatr}) and (\ref{def_ML_H}).
\end{theorem}

Theorem 2 is essentially a matrix generalization of Theorem 1, or more precisely of Proposition 1, which improves the estimates as seen in Sec.~\ref{SC_Dmu}. The constants $\overline{M}$, $\lambda$, $M_{\delta}$ and $\lambda_{\delta}$ of Eq.~\eqref{def_MLd} are respectively replaced by two-dimensional vectors $\overline{{\mathbf M}}$, ${\mathbf M_\delta}$, and $2\times 2$ matrices ${\mathbb \Lambda}$, ${\mathbb \Lambda}_\delta$.

The proof of Theorem 2 follows the same lines as that of Theorem 1. We write
\begin{eqnarray*}
\begin{pmatrix}
\Vert {\mathbf r}(t) - \overline{\mathbf r}(t)\Vert \\
\Vert {\mathbf p}(t) - \overline{\mathbf p}(t)\Vert 
\end{pmatrix}  
= \varepsilon \begin{pmatrix}
\Vert \int_0^t [ {\mathbf \nabla}_{\mathbf p} H(\tau,  {\mathbf z}(\tau), \varepsilon) - {\mathbf \nabla}_{\mathbf p} \langle H\rangle(\tau,  {\mathbf u}(\tau), \varepsilon) ] {\rm d}\tau \Vert \\
\Vert \int_0^t [ {\mathbf \nabla}_{\mathbf r} H(\tau,  {\mathbf z}(\tau), \varepsilon) - {\mathbf \nabla}_{\mathbf r} \langle H\rangle(\tau,  {\mathbf u}(\tau), \varepsilon) ] {\rm d}\tau \Vert 
\end{pmatrix} \\
\le \varepsilon \begin{pmatrix}
\int_0^t \Vert {\mathbf \nabla}_{\mathbf p} H(\tau,  {\mathbf z}(\tau), \varepsilon) - {\mathbf \nabla}_{\mathbf p} H(\tau,  {\mathbf u}(\tau), \varepsilon) \Vert {\rm d}\tau  + 
\Vert \int_0^t {\mathbf \nabla}_{\mathbf p}\Delta H(\tau,  {\mathbf u}(\tau), \varepsilon) {\rm d}\tau \Vert \\
\int_0^t \Vert  {\mathbf \nabla}_{\mathbf r} H(\tau,  {\mathbf z}(\tau), \varepsilon) - {\mathbf \nabla}_{\mathbf r} H(\tau,  {\mathbf u}(\tau), \varepsilon) \Vert {\rm d}\tau +
\Vert \int_0^t {\mathbf \nabla}_{\mathbf r}\Delta H(\tau,  {\mathbf u}(\tau), \varepsilon) {\rm d}\tau \Vert 
\end{pmatrix} \\
\le \varepsilon {\mathbb \Lambda} \begin{pmatrix}
\int_0^t \Vert {\mathbf r}(\tau) - \overline{\mathbf r}(\tau)\Vert {\rm d}\tau \\
\int_0^t \Vert {\mathbf p}(\tau) - \overline{\mathbf p}(\tau)\Vert {\rm d}\tau
\end{pmatrix} + \varepsilon {\mathbf C} \,,
\end{eqnarray*}
where ${\mathbf C}$ is defined by
$$ 
{\mathbf C} = \sup_{t\in [0,L/\varepsilon]} 
\begin{pmatrix}
\Vert \int_0^t {\mathbf \nabla}_{\mathbf p}\Delta H(\tau,  {\mathbf u}(\tau), \varepsilon) {\rm d}\tau \Vert \\
\Vert \int_0^t {\mathbf \nabla}_{\mathbf r}\Delta H(\tau,  {\mathbf u}(\tau), \varepsilon) {\rm d}\tau \Vert 
\end{pmatrix} \,.
$$

We now use the following generalization of Gr{\"o}nwall's inequality: 
\begin{lemma}[\cite{Chandra1976}]
Let the vector ${\mathbf a}(t)$ and the nonnegative matrices ${\mathbb G}(t)$ and ${\mathbb H}(t)$ be functions of the scalar variable $t$. Assume that ${\mathbb H}(t) {\mathbb G}(t)$ and $\int_{t_0}^t {\mathbb H}(s) {\mathbb G}(s) {\rm d}s$ commute for $t> t_0$. If the inequality
$$
{\bm \sigma}(t) \le {\mathbf a}(t) + {\mathbb G}(t) \int_{t_0}^t {\mathbb H}(\tau) {\bm \sigma}(\tau) {\rm d}\tau 
$$
holds for all $t>t_0$, then
$$
{\bm\sigma}(t) \le {\mathbf a}(t) + {\mathbb G}(t) \int_{t_0}^t \exp{\left( \int_{\tau}^t {\mathbb H}(s) {\mathbb G}(s) {\rm d}s \right)} {\mathbb H}(\tau) {\mathbf a}(\tau) {\rm d}\tau 
$$
holds for $t>t_0$.
\end{lemma}

In our case ${\mathbf a}(t) = \varepsilon {\mathbf C}$, ${\mathbb G}(t) = \varepsilon {\mathbb \Lambda}$ independent of $t$, and ${\mathbb H}(t)$ is the identity matrix. This leads to:
\begin{eqnarray*}
\begin{pmatrix}
\Vert {\mathbf r}(t) - \overline{\mathbf r}(t)\Vert \\
\Vert {\mathbf p}(t) - \overline{\mathbf p}(t)\Vert 
\end{pmatrix} \le \left[ \mathbb{I}_2 + \varepsilon {\mathbb \Lambda} \int_0^t {\rm d}\tau\, {\rm e}^{(t-\tau) \varepsilon {\mathbb \Lambda} } \right] \varepsilon {\mathbf C} = {\rm e}^{\varepsilon t {\mathbb \Lambda}} \varepsilon {\mathbf C} \,.
\end{eqnarray*}
Here we clearly notice that the estimates for ${\bf C}$ are obtained from the trajectories ${\bf u}(t)$ in the averaged system.

We evaluate the vector ${\mathbf C}$ in the case where $H(t,  {\mathbf z}, \varepsilon)$ is periodic in $t$, i.e., $H(t,  {\mathbf z}, \varepsilon) = H(t+T,  {\mathbf z}, \varepsilon)$ for all $t$. Proceeding in the same way as in the proof of Theorem 1, we partition the interval $[0,t]$: For $n T\le t < (n+1)T$, i.e., $n=\lfloor t/T\rfloor$, we get (say, for the first component):
\begin{eqnarray*}
\left\Vert \int_0^t {\mathbf \nabla}_{\mathbf p}\Delta H(\tau,  {\mathbf u}(\tau), \varepsilon) {\rm d}\tau \right\Vert \le
\sum_{i=1}^n  \int_{(i-1)T}^{iT} \Vert {\mathbf \nabla}_{\mathbf p}\Delta H(\tau,  {\mathbf u}(\tau), \varepsilon) - {\mathbf \nabla}_{\mathbf p}\Delta H(\tau,  {\mathbf u}_i, \varepsilon) \Vert {\rm d}\tau \\
+ \int_{nT}^t \Vert {\mathbf \nabla}_{\mathbf p}\Delta H(\tau,  {\mathbf u}(\tau), \varepsilon) \Vert \, {\rm d}\tau \,,
\end{eqnarray*}
where ${\mathbf u}_i = {\mathbf u}((i-1)T)\,, \, i=1,\dots,n$ denote the values of ${\mathbf u}$ at each period. It follows that
$$
\left\Vert \int_0^t {\mathbf \nabla}_{\mathbf p}\Delta H(\tau,  {\mathbf u}(\tau), \varepsilon) {\rm d}\tau \right\Vert \le \sum_{i=1}^n \int_{(i-1)T}^{iT} [ \lambda^\delta_{rp} \Vert \overline{\mathbf r}(\tau) - \overline{\mathbf r}_i \Vert + \lambda^\delta_{pp} \Vert \overline{\mathbf p}(\tau) - \overline{\mathbf p}_i \Vert ] {\rm d}\tau + M^\delta_p T \,.
$$
Using the fact that for $\tau\in[(i-1)T,iT]$
$$
{\mathbf u}(\tau) = {\mathbf u}_i + \int_{(i-1)T}^{\tau}  \varepsilon\, \langle {\mathbf Z}\rangle({\mathbf u}(s), \varepsilon) \, {\rm d}s\,,
$$
we get
$$
\Vert \overline{\mathbf r}(\tau) - \overline{\mathbf r}_i \Vert \le \varepsilon M_p (\tau - (i-1)T) \,, \qquad
\Vert \overline{\mathbf p}(\tau) - \overline{\mathbf p}_i \Vert \le \varepsilon M_r (\tau - (i-1)T) \,,
$$
so that
\begin{eqnarray*}
\left\Vert \int_0^t {\mathbf \nabla}_{\mathbf p}\Delta H(\tau,  {\mathbf u}(\tau), \varepsilon) {\rm d}\tau \right\Vert &\le&
\varepsilon n \frac{T^2}{2} (\lambda^\delta_{rp} M_p + \lambda^\delta_{pp} M_r) + M^\delta_p T, \\
&\le&
T \left[\frac{\varepsilon t}{2} (\lambda^\delta_{rp} M_p + \lambda^\delta_{pp} M_r) + M^\delta_p \right] \, ,
\end{eqnarray*}
for all $t\leq L/\varepsilon$. 
Performing the same estimate for the second component leads to
$$
{\mathbf C} = T \left[\frac{L}{2} {\mathbb \Lambda}_\delta \overline{\mathbf M} + {\mathbf M_\delta} \right] \,,
$$
from which Eq.~\eqref{thBog_H} follows.

\noindent {\it Remark :} Thanks to the fact that the elements on the diagonal of ${\mathbb \Lambda}$ are identical (which comes from the Hamiltonian character of the dynamics) the exponential ${\rm e}^{ {\mathbb \Lambda} L}$ has a closed expression. For example, in the usual case where Hamiltonian $H$ is at most quadratic in momenta (i.e., $\lambda_{pp} = 1$) we have:
\begin{equation} \label{expLambda}
{\rm e}^{ {\mathbb \Lambda} L} = {\rm e}^{\lambda_{rp} L}
\begin{pmatrix}
\cosh(\sqrt{\lambda_{rr}} L) & \quad \frac{1}{\sqrt{\lambda_{rr}}} \sinh(\sqrt{\lambda_{rr}} L)
\\ \\  \sqrt{\lambda_{rr}} \sinh(\sqrt{\lambda_{rr}} L) & \quad \cosh(\sqrt{\lambda_{rr}} L)
\end{pmatrix} \,.
\end{equation}

Equation~\eqref{thBog_H} is the matrix equivalent of the scalar expression \eqref{thBogd} of Proposition 1. The advantage of the matrix formulation is evident in the treatment of the simple example (\ref{sflow1}) considered in section \ref{sec_disc}. We have in this case:
$$
{\mathbb \Lambda} = 
\begin{pmatrix}
0 & 1 \\ 0 & 0
\end{pmatrix} \,, \qquad
\overline{\mathbf M} = 
\begin{pmatrix}
M_p \\ 0
\end{pmatrix} \,, \qquad
{\mathbb \Lambda}_\delta = {\mathbb 0} \,, \qquad
{\mathbf M_\delta} = 
\begin{pmatrix}
0 \\ 1
\end{pmatrix} \,.
$$
In this example, we notice that the value of $M_p$ is irrelevant since $\overline{\mathbf{M}}$ is multiplied by the null matrix.
Theorem 2 gives, for all $t\leq t_{\rm max}$:
$$
| x(t) - \overline{x}(t) | \le 2\pi \varepsilon^2 t_{\rm max} \,, \quad | p(t) - \overline{p}(t) | \le 2\pi \varepsilon \,,
$$
to be compared with the estimate coming from the exact solution:
$$
| x(t) - \overline{x}(t) | = \varepsilon^2 |1-\cos t | \le 2\varepsilon^2 \,, \quad | p(t) - \overline{p}(t) | = \varepsilon |\sin t | \le \varepsilon \,.
$$

As it has been pointed out in section \ref{sec_disc}, Theorem 1 gives: 
$$
| x(t) - \overline{x}(t) | + | p(t) - \overline{p}(t) | \leq 2\pi \varepsilon  M (\varepsilon t_{\rm max} + 2) {\rm e}^{\varepsilon t_{\rm max}} \,, 
$$
and the modified version of Proposition 1 (with $\lambda_\delta=0$, $M_\delta = 1$) leads to :
$$
| x(t) - \overline{x}(t) | + | p(t) - \overline{p}(t) | \leq 2\pi \varepsilon  M {\rm e}^{\varepsilon t_{\rm max}} \,,
$$
for all $\vert p\vert \leq M-1$ and $t\leq t_{\rm max}$. We notice that the matrix formulation in Theorem 2 allowed the removal of the exponentially divergent term for this specific case, since ${\rm e}^{t {\mathbb \Lambda}}$ is not necessarily exponential in time when ${\mathbb \Lambda}$ is a matrix. Next, we consider the running example of a one-dimensional soft-Coulomb potential to figure out if Theorem 2 improves the estimates and allows us to conclude on the validity of the Kramers-Henneberger approximation.  

\subsection{Application to a one-dimensional soft-Coulomb potential}
For a one-dimensional soft-Coulomb potential, the matrices and vectors appearing in Theorem 2 are given by:
$$
{\mathbb \Lambda} = 
\begin{pmatrix}
0 & 1 \\ B_{W,2} & 0
\end{pmatrix} \,, \quad
\overline{\mathbf M} = 
\begin{pmatrix}
B_p \\ B_{W,1}
\end{pmatrix} \,, \quad
{\mathbb \Lambda}_\delta = 
\begin{pmatrix}
0 & 0 \\ B_{W,2} & 0
\end{pmatrix} \,, \quad
{\mathbf M_\delta} = 
\begin{pmatrix}
0 \\ B_{W,1}
\end{pmatrix} \, ,
$$
which gives, using Eqs.~\eqref{thBog_H}-\eqref{expLambda}:
\begin{equation} \label{eqn:sigma2H}
\begin{pmatrix}
| \chi(\phi) - \overline{\chi}(\phi)  | 
\\
| \xi(\phi)  - \overline{\xi}(\phi) | 
\end{pmatrix} \le 2\pi \varepsilon \left( \frac{L}{2} B_p B_{W,2} + B_{W,1} \right)
\begin{pmatrix}
\frac{1}{ \sqrt{ B_{W,2} } } \sinh(\sqrt{B_{W,2}} L)
\\
\cosh(\sqrt{B_{W,2}} L)
\end{pmatrix} .
\end{equation}
Going back to the original variables $x_{\rm e}$, $p_{\rm e}$ and $t$, and taking $E_0/\omega$ as the rescaling factor for momenta, we have
\begin{eqnarray} \label{eqn:sigma2X}
&& \frac{1}{q} \, \left| x_{\rm e}(t) - \overline{x}_{\rm e}(t) \right| \le \frac{T}{q^{3/2}} \, \left( \frac{t_{\rm max}}{2 q^{3/2}} B_p B_{W,2} + B_{W,1}\right) \frac{1}{ \sqrt{ B_{W,2} } } \sinh\left(\frac{t_{\rm max} \sqrt{B_{W,2}} }{q^{3/2}} \right) ,
\\ \label{eqn:sigma2P}
&& \frac{\omega}{E_0} \, \left| p_{\rm e}(t) - \overline{p}_{\rm e}(t) \right| \le \frac{\varepsilon T}{q^{3/2}} \, \left( \frac{t_{\rm max}}{2 q^{3/2}} B_p B_{W,2} + B_{W,1}\right) \cosh\left(\frac{t_{\rm max} \sqrt{B_{W,2}} }{q^{3/2}} \right).
\end{eqnarray} 
We first notice that there is still an exponential behavior in time for $t_{\rm max}$ large, but the exponent is $\sqrt{B_{W,2}}$ compared with the exponent for Theorem 1 and Proposition 1, $\max(1,B_{W,2})$. This reduction of exponent is crucial for small $\mu$s, and is one of the main ingredients for the improvement of the estimates.  
In Fig.~\ref{fig:tradi}, we have plotted the bounds $\sigma$ for $\left| x_{\rm e}(t) - \overline{x}_{\rm e} (t) \right|/q + \sqrt{q} \left| p_{\rm e}(t) - \overline{p}_{\rm e}(t) \right|$ where the two contributions are obtained using Eqs.~\eqref{eqn:sigma2X}-\eqref{eqn:sigma2P}. We notice that for large $\mu$ the estimates obtained from Theorem 2 appear to be equivalent to the ones of Proposition 1. The main difference manifests itself for small values of $\mu$, where the estimates improve closer to the ionic core. 

\begin{figure}
    \centering
    \includegraphics[width=0.8\textwidth]{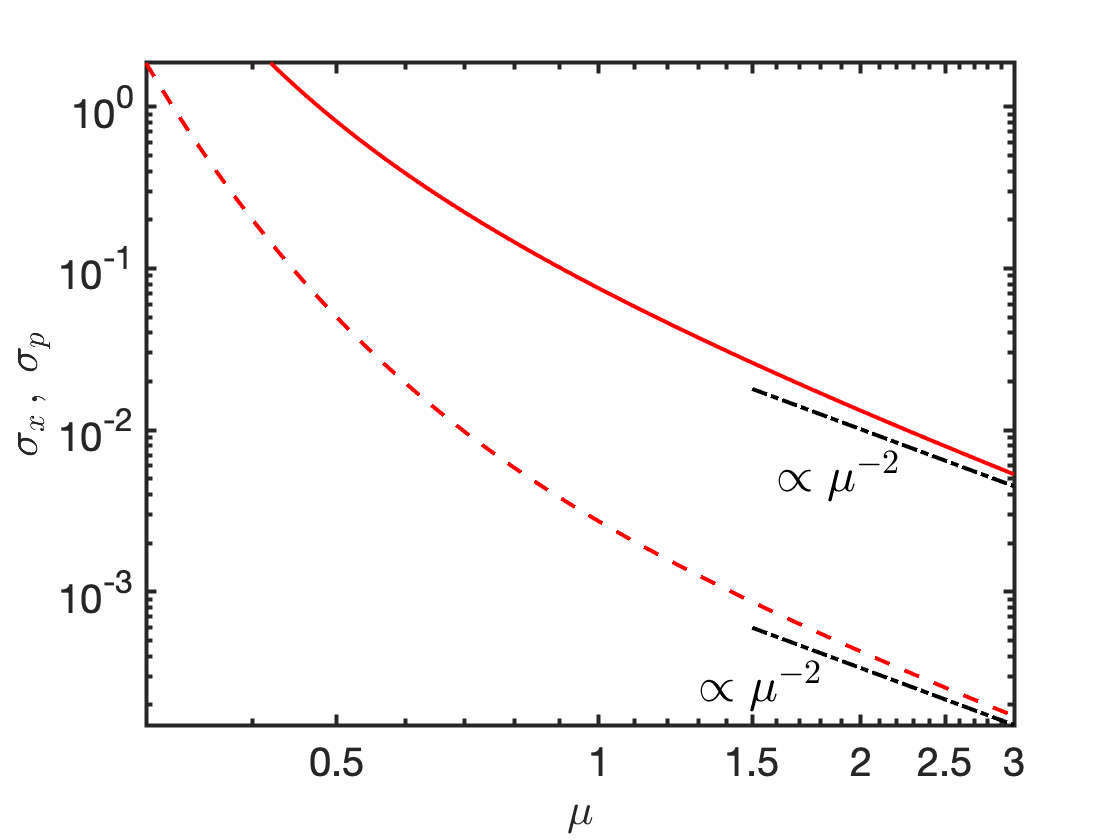}
    \caption{
Loglog plot of $\sigma_{x} = \frac{\omega^2}{E_0} \, \left| x_{\rm e}(t) - \overline{x}_{\rm e} (t) \right|$ (red solid line) and $\sigma_{p} = \frac{\omega}{E_0} \, \left| p_{\rm e}(t) - \overline{p}_{\rm e}(t) \right|$ (red dash-dotted line) given by equations Eqs.~\eqref{eqn:sigma2X}-\eqref{eqn:sigma2P} as a function of $\mu$ for $I = 5\times 10^{15}$ W cm$^{-2}$, $\omega = 0.0584$, $a=1$ and $t_{\rm max} = 5$ laser cycles.
}
    \label{fig:sigmaXP}
\end{figure}

In order to be more quantitative and compare the results of Theorem 1, Proposition 1 and Theorem 2 for the one-dimensional soft-Coulomb potential, we compute the estimates for large and small distances of the electron trajectory from the ionic core. This allows us to better understand the different behaviours displayed on Fig.~\ref{fig:tradi} and identify quantitatively the improvement of each estimate. For each case, we give the leading term in the expression
$$
\sigma = \max_{t\leq t_{\rm max}} \left( \frac{1}{q} \, \left| x_{\rm e}(t) - \overline{x}_{\rm e}(t) \right| + \sqrt q \, \left| p_{\rm e}(t) - \overline{p}_{\rm e}(t) \right| \right)\,.
$$

\noindent Large distance: $\mu\gg 1$.
\begin{eqnarray} 
&& 
\sigma^{\rm Th 1} \simeq \frac{T}{q^{3/2}} \, \mathrm{e}^{t_{\rm max}\, q^{-3/2} } \left( \frac{t_{\rm max}}{q^{3/2}} + 2 \right) \sqrt{\frac{2}{\mu}} \,, \nonumber \\
&& \label{eqn:sigma2LargeProp1}
\sigma^{\rm Prop 1} \simeq \frac{T}{q^{3/2}} \, \mathrm{e}^{t_{\rm max}\, q^{-3/2} } \frac{1}{\mu^2} \,, \\
&& \label{eqn:sigma2LargeTh2}
\sigma^{\rm Th 2} \simeq \frac{T}{q^{3/2}} \, \left( 1 +  \frac{t_{\rm max}}{q^{3/2}} \right) \frac{1}{\mu^2} \,,
\end{eqnarray}
where estimates \eqref{eqn:sigma2LargeProp1}-\eqref{eqn:sigma2LargeTh2} are valid for $\mu$ such that
$$
\frac{\sqrt{2} \,t_{\rm max}}{(\mu q)^{3/2}} \ll 1 \,.
$$
We see that the estimates from Proposition 1 and Theorem 2 decrease faster with distance than the estimate relative to Theorem 1. Moreover, Theorem 2 gives a better result than Proposition 1 if the pulse duration $t_{\rm max}$ is larger than $q^{3/2}$.

\bigskip

\noindent Small  distance: $\alpha \ll \mu\ll 1$. 
\begin{eqnarray*}
&& 
\sigma^{\rm Th 1} \simeq \frac{T t_{\rm max}}{q^{3}} \, \mathrm{e}^{2 t_{\rm max}\, q^{-3/2} \mu^{-3} } \frac{2}{\mu^5} \,, \\
&& 
\sigma^{\rm Prop 1} \simeq \frac{T t_{\rm max}}{q^{3}} \, \mathrm{e}^{2 t_{\rm max}\, q^{-3/2} \mu^{-3} } \frac{1}{\mu^5} \,, \\
&& 
\sigma^{\rm Th 2} \simeq \frac{T t_{\rm max}}{q^{3}} \, \mathrm{e}^{\sqrt{2} t_{\rm max}\,q^{-3/2} \mu^{-3/2}  }  \frac{\sqrt{2}}{\mu^{7/2}} \,,
\end{eqnarray*}
where estimates are valid for $\mu$ such that
$$
\frac{\sqrt{2} \,t_{\rm max}}{(\mu q)^{3/2}} \gg 1 \,.
$$
We see that the estimate from Theorem 2 diverges slower than those coming from Theorem 1 and Proposition 1 as $\mu$ goes to zero. This is in particular due to the reduction of the exponent in the exponential, from $\mu^{-3}$ to $\mu^{-3/2}$. As a consequence, in the two ranges of $\mu$, Theorem 2 provides significantly much better estimates than Theorem 1 and Proposition 1. 

One additional advantage with the formulation of Theorem 2 is that it gives separate bounds on the positions and the momenta. These bounds are represented in Fig.~\ref{fig:sigmaXP} where the positions are rescaled by $E_0/\omega^2$ and the momenta by $E_0/\omega$. Using these scalings, we see that the dominant error as a result of the KH approximation comes from the positions getting close to the ionic core. 

\subsection{Discussion}

As we have seen above, the amended version of Bogolyubov's theorem provides better estimates than the original version. The main reason is that this version is more suited to Hamiltonian systems. Does it allow us to conclude on the validity of the Kramers-Henneberger approximation? As discussed in Sec.~\ref{sec_disc}, the answer to the validity of the KH approximation still depends on the values of the parameters and on the region of phase space in which the trajectory evolves. If the trajectory is sufficiently far from the ionic core (at least, at a distance of one quiver radius), the approximation is valid. The validity gets better as the trajectory gets further away from the core. Using a more apt version of the Bogolyubov's averaging theorem adapted to Hamiltonian flows, the validity is confirmed closer to the ionic core than it was the case with the original version of the theorem. However, all versions of the averaging theorem lead to the same conclusions if $\mu$ is sufficiently close to zero. This means that when the trajectory is allowed to cross $x=0$, the bounds between this trajectory and the one of the averaged system become large. 

A numerical investigation of $H$ and $\langle H \rangle$ confirms that the two Cauchy problems do not separate exponentially as predicted by averaging theorems. The behavior is much more complex, depending on the region of phase space visited by the trajectories. In order to illustrate this complex behavior a Matlab~\cite{MATLAB:2021} script \texttt{KHBogolyubov.mlx} is available at \texttt{github.com/cchandre/KH}. This script computes and displays the trajectories in phase space for both Hamiltonians $H$ and $\langle H\rangle$ starting with the same initial conditions, as well as their distance as a function of time, in order to compare it with the results obtained using Theorem 1, Proposition 1 and Theorem 2. This script also contains the codes to reproduce Figs.~\ref{fig:tradi} and \ref{fig:sigmaXP}. 

One of the main arguments of the KH approximation is the possibility to create stable laser-dressed states located near the minimum of the KH potential. However, this minimum is always at a distance smaller than one quiver radius. The estimates obtained by Bogolyubov's theorem or their amended versions do not work in this region of phase space, mostly regardless of the parameters of the laser field. As a consequence, we cannot conclude on the validity of the KH approximation close to the local minimum of the KH potential which is at the core of the KH atom.  

\section*{Acknowledgments}
CC and JD acknowledge the funding from the European Union's Horizon 2020 research and innovation program under the Marie Sk\l odowska-Curie Grant Agreement No. 734557.

%\bibliographystyle{elsarticle-num} 
%\bibliography{biblio}

\begin{thebibliography}{10}
\expandafter\ifx\csname url\endcsname\relax
  \def\url#1{\texttt{#1}}\fi
\expandafter\ifx\csname urlprefix\endcsname\relax\def\urlprefix{URL }\fi
\expandafter\ifx\csname href\endcsname\relax
  \def\href#1#2{#2} \def\path#1{#1}\fi

\bibitem{vandeSand1999}
G.~van~de Sand, J.~M. Rost, Irregular orbits generate higher harmonics, Phys.
  Rev. Lett. 83 (1999) 524.

\bibitem{Corkum1993}
P.~B. Corkum, Plasma perspective on strong field multiphoton ionization, Phys.
  Rev. Lett. 71 (1993) 1994.

\bibitem{Schafer1993}
K.~J. Schafer, B.~Yang, L.~F. DiMauro, K.~C. Kulander, Above threshold
  ionization beyond the high harmonic cutoff, Phys. Rev. Lett. 70 (1993) 1599.

\bibitem{Becker2008_ContP}
W.~Becker, H.~Rottke, Many-electron strong-field physics, Cont. Phys. 49 (2008)
  199.

\bibitem{Grobe1991}
R.~Grobe, C.~K. Law, Stabilization in superintense fields: A classical
  interpretation, Phys. Rev. A 44 (1991) R4114.

\bibitem{Gavrila2002}
M.~Gavrila, Atomic stabilization in superintense laser fields, J. Phys. B: At.
  Mol. Opt. Phys. 35 (2002) R147.

\bibitem{Norman2015}
M.~J. Norman, C.~Chandre, T.~Uzer, P.~Wang, Nonlinear dynamics of ionization
  stabilization of atoms in intense laser fields, Phys. Rev. A 91 (2015)
  023406.

\bibitem{Kramers1956}
H.~A. Kramers, Collected Scientific Papers, North-Holland Publishing Company,
  Amsterdam, 1956.

\bibitem{Henneberger1968}
W.~C. Henneberger, Perturbation method for atoms in intense light beams, Phys.
  Rev. Lett. 21 (1968) 838.

\bibitem{Breuer1992}
H.~Breuer, K.~Dietz, M.~Holthaus, A remark on the {K}ramers-{H}enneberger
  transformation, Phys. Lett. A 165 (1992) 341.

\bibitem{Reed1993}
V.~C. Reed, K.~Burnett, P.~L. Knight, Harmonic generation in the
  {K}ramers-{H}enneberger stabilization regime, Phys. Rev. A 47 (1993) R34.

\bibitem{Volkova1994}
E.~A. Volkova, A.~M. Popov, O.~V. Smirvova, Stabilization of atoms in a strong
  field and the {K}ramers-{H}enneberger approximation, JETP 79 (1994) 736.

\bibitem{Popov1999}
A.~M. Popov, O.~V. Tikhonova, E.~A. Volkova, Applicability of the
  {K}ramers-{H}enneberger approximation in the theory of strong-field
  ionization, J. Phys. B: At. Mol. Opt. Phys. 32 (1999) 3331.

\bibitem{Forre2005}
M.~F\o{}rre, S.~Selst\o{}, J.~P. Hansen, L.~B. Madsen, Exact nondipole
  {K}ramers-{H}enneberger form of the light-atom hamiltonian: An application to
  atomic stabilization and photoelectron energy spectra, Phys. Rev. Lett 95
  (2005) 043601.

\bibitem{Ivanov2005}
I.~A. Ivanov, A.~S. Kheifets, On the use of the {K}ramers–{H}enneberger
  {H}amiltonian in multi-photon ionization calculations, J. Phys. B: At. Mol.
  Opt. Phys. 38 (2005) 2245.

\bibitem{He2020}
P.~L. He, Z.~H. Zhang, F.~He, Young’s double-slit interference in a hydrogen
  atom, Phys. Rev. Lett. 124 (2020) 163201.

\bibitem{Morales2011}
F.~Morales, M.~Richter, S.~Patchkovskii, O.~V. Smirnova, Imaging the
  {K}ramers–{H}enneberger atom, PNAS 108 (2011) 16906.

\bibitem{Richter2013}
M.~Richter, F.~Patchkovskii, F.~Morales, O.~Smirnova, M.~Ivanov, The role of
  the {K}ramers–{H}enneberger atom in the higher-order {K}err effect, New J.
  Phys. 15 (2013) 083012.

\bibitem{Wei2017}
Q.~Wei, P.~Wang, S.~Kais, D.~Herschbach, Pursuit of the {K}ramers-{H}enneberger
  atom, Chem. Phys. Lett. 683 (2017) 240.

\bibitem{Smirnova2000}
O.~V. Smirnova, Validity of the {K}ramers–{H}enneberger approximation, JETP
  90 (2000) 609.

\bibitem{Bogolyubov1961}
N.~Bogolyubov, Y.~Mitropol’skii, Asymptotic Methods in the Theory of
  Non-Linear Oscillations, Gordon and Breach, New York, 1961.

\bibitem{Zhuravlev1988}
V.~F. Zhuravlev, D.~M. Klimov, Applied Methods in the Vibration Theory, Nauka,
  Moscow, 1988, (in Russian).

\bibitem{Mitropolskii1997}
Y.~A. Mitropolsky, N.~V. Dao, Applied Asymptotic Methods in Nonlinear
  Oscillations, Springer Netherlands, 1997.

\bibitem{Dubois2018}
J.~Dubois, S.~A. Berman, C.~Chandre, T.~Uzer, Capturing photoelectron motion
  with guiding centers, Phys. Rev. Lett. 121 (2018) 113202.

\bibitem{Dubois2018_PRE}
J.~Dubois, S.~A. Berman, C.~Chandre, T.~Uzer, Guiding-center motion for
  electrons in strong laser fields, Phys. Rev. E 98 (2018) 052219.

\bibitem{Chandra1976}
J.~Chandra, P.~W. Davis, Linear generalizations of {G}r{\"o}nwall's inequality,
  Proc. Amer. Math. Soc. 60 (1976) 157.

\bibitem{MATLAB:2021}
MATLAB, version 9.10.0 (R2021a), The MathWorks Inc., Natick, Massachusetts,
  2021.

\end{thebibliography}

\end{document}